# Langmuir-Blodgett Assembly of Densely Aligned Single-Walled Carbon Nanotubes from Bulk Materials

Xiaolin Li, Li Zhang, Xinran Wang, Iwao Shimoyama, Xiaoming Sun, Won-Seok Seo, Hongjie Dai*

*Department of Chemistry, Stanford University, Stanford, CA 94305, USA.*

hdai@stanford.edu

Single-walled carbon nanotubes (SWNTs) exhibit advanced properties desirable for high performance nanoelectronics. Important to future manufacturing of high-current, speed and density nanotube circuits is large-scale assembly of SWNTs into densely aligned forms.[1] Despite progress in oriented synthesis and assembly including the Langmuir-Blodgett (LB) method,[2-9] no method exists for producing assemblies of pristine SWNTs (free of extensive covalent modifications) with both high density and high degree of alignment of SWNTs. Here, we develop a LB method achieving monolayers of aligned non-covalently functionalized SWNTs from organic solvent with dense packing. The monolayer SWNTs are readily patterned for device integration by microfabrication, enabling the high currents (~3mA) SWNT devices with narrow channel widths. Our method is generic for different bulk materials with various diameters.

Suspensions of as-grown laser-ablation and Hipco SWNTs in 1,2-dichloroethane (DCE) solutions of poly(m-phenylenevinylene -co-2,5-dioctoxy-p-phenylenevinylene) (PmPV) were prepared by sonication, ultra centrifugation and filtration (see supplementary information). The suspension contained mostly individual nanotubes (average diameter~1.3nm and ~1.8nm respectively for Hipco and laser-ablation materials, mean length ~500nm, Fig.1d and 1e) well solubilized in DCE without free unbound PmPV. PmPV is known to exhibit high binding affinity to SWNT sidewall via $\pi$ stacking of its conjugated backbone (Fig.1a) and thus impart solubility of nanotubes in organic solvents.[10] Indeed, we obtained homogeneous suspensions of nanotubes in PmPV solutions. However, we found that DCE was the only solvent in which PmPV bound SWNTs remained stably suspended when free unbound PmPV molecules were removed (Inset of Fig.1b). The PmPV treated SWNTs exhibited no aggregation in DCE over several months. DCE without PmPV could suspend low concentrations of SWNTs (~50X lower than with PmPV functionalization), insufficient for LB formation, especially for larger SWNTs in laser materials with lower solubility.

The excitation and emission spectra of PmPV bound SWNTs (in PmPV-SWNT solution with excess PmPV removed) exhibited ~20nm and ~3nm shifts respectively relative to those of pure PmPV in DCE (Fig.1b), providing spectroscopic evidence of strong interaction between PmPV and SWNTs. No change in the spectra was observed with the highly stable PmPV-SWNT/DCE suspension for months, indicating strong binding of PmPV on SWNT without detachment in DCE. The fact that PmPV-SWNTs were not stably suspended in other solvents without excess PmPV and that addition of large amounts of these solvents (e.g., chloroform) into a PmPV-SWNT/DCE suspension causing nanotube precipitation suggested significant detachment of PmPV from nanotubes in most organic solvents. The unique stability of PmPV coating on SWNT in DCE over other solvents is not fully understood currently. Nevertheless, it is highly desirable for chemical assembly of high quality nanotubes and integrated devices since it enables non-covalently functionalized SWNTs (both large diameter laser and small diameter Hipco materials) soluble in organics in nearly pristine form, as gleaned from the characteristic UV-vis-NIR absorbance (Fig.1c) and Raman signatures of non-covalently modified SWNTs (Fig.2c).

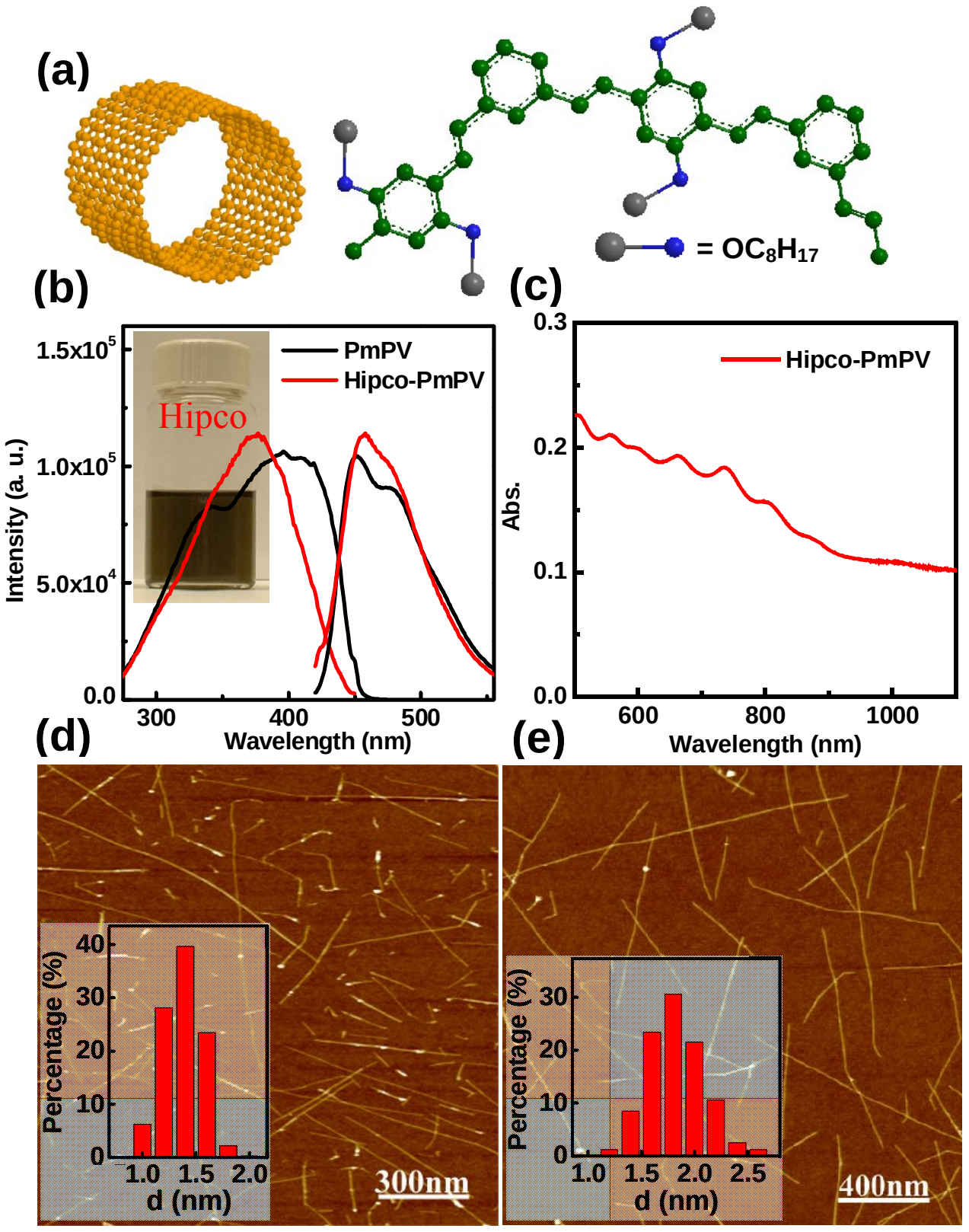

***Figure 1.*** PmPV functionalized SWNTs. (a) Schematic drawings of a SWNT and two units of a PmPV chain. (b) Excitation and fluorescence spectra of pure PmPV in DCE vs. PmPV bound Hipco SWNTs in DCE. Inset: photograph of PmPV coated Hipco SWNTs suspended in DCE without excess PmPV in the solution. (c) UV-vis-NIR spectrum of PmPV suspended Hipco SWNTs with no excess PMPV. (d) & (e) Atomic force microscopy (AFM) images of Hipco and laser-ablation SWNTs randomly deposited on a substrate from solution. Insets: Diameter distributions.

PmPV-SWNTs were spread on a water subphase from a DCE solution, compressed upon DCE vaporization to form a LB film using compression-retraction-compression cycles to reduce hysteresis (supplementary Fig.S1, S2&S3) and then vertically transferred onto a $SiO_2$ or any other substrate (glass, plastic, etc.). Organic solutions of stably suspended SWNTs without excess free polymer are critical to high density SWNT LB film formation. Microscopy (Fig.2a&2b) and spectroscopy (Fig.2c&2d) characterization revealed high quality densely aligned SWNTs (normal to the compression and substrate pulling direction) formed uniformly over large substrates for both Hipco and laser ablation materials. Height of the film relative to tube-free regions of the substrate was <2nm under AFM, suggesting monolayer of packed SWNTs. Micro-Raman spectra of the SWNTs showed ~ $\cos^2\alpha$ polarization dependence of the G band (~1590cm$^{-1}$) intensity (Fig. 2d), where α is the angle between the laser polarization and the SWNT alignment direction. The peak to valley ratio of the Raman intensities was ~8 with little variation over the substrate, indicating alignment of SWNTs over large areas. Nevertheless, imperfections existed in the quasi-aligned dense SWNT assembly including voids, bending and looping of nanotubes formed during the compression process for LB film formation due to the high aspect ratio (diameter <~2nm, length ~200nm-1μm) and mechanical flexibility of SWNTs.

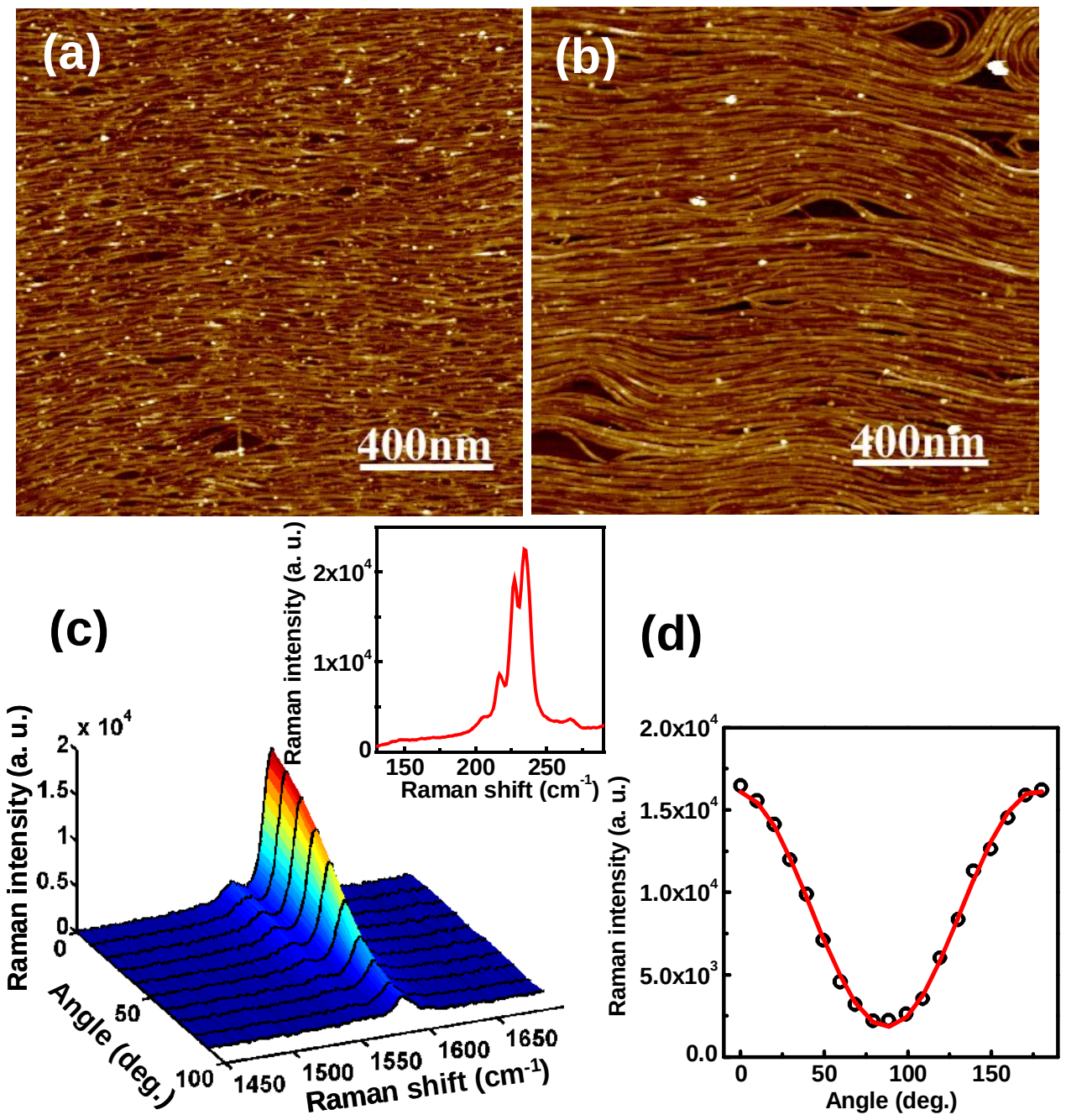

***Figure 2.*** LB monolayers of aligned SWNTs. (a) AFM image of a LB film of Hipco SWNTs on a $SiO_2$ substrate. (b) AFM image of a LB film of laser-ablation SWNTs. (c) Raman spectra of the G line of a Hipco SWNT LB film recorded at various angles (α) between the polarization of laser excitation and SWNT alignment direction. Inset: Raman spectrum showing the radial breathing mode (RBM) region of the Hipco LB film at α~0. (d) G line (1590cm$^{-1}$) intensity vs. angle α for the Hipco SWNT LB film in (c). The red curve is a $\cos^2\alpha$ fit.

Our aligned SWNT monolayers on oxide substrates can be treated as carbon-nanotube on insulator (CNT_OI) materials for patterning and integration into potential devices, much like how Si on insulator (SOI) has been used for electronics. We used lithographic patterning techniques and oxygen plasma etching to remove unwanted nanotubes and form patterned arrays of squares or rectangles comprised of aligned SWNTs (Fig.3a and 3b). We then fabricated arrays of two-terminal devices with Ti/Au metal source (S) and drain (D) contacting massively parallel SWNTs in ~10 μm wide S-D regions with channel length ~250nm (Fig.3c and 3d). Current vs. bias voltage (I-V) measurements showed that such devices made from Hipco SWNTs were more than 25 times more resistive than similar devices made from laser-ablation SWNTs, with currents reaching ~0.13mA and ~3.5mA respectively at a bias of 3 V through collective current carrying of SWNTs in parallel (Fig.3e and 3f). Further, Hipco SWNT devices exhibited higher non-linearity in the I-V characteristics than laser ablation nanotubes (Fig.3e). These results were attributed to the diameter difference between Hipco and laser-ablation materials. Hipco SWNTs were small in diameter with many tubes ≤1.2nm, giving rise to high (non-ohmic) contact resistance for both semiconducting and metallic SWNTs.[11] Smaller SWNTs could also be more susceptible to defects and disorder, contributing to degraded current carrying ability.

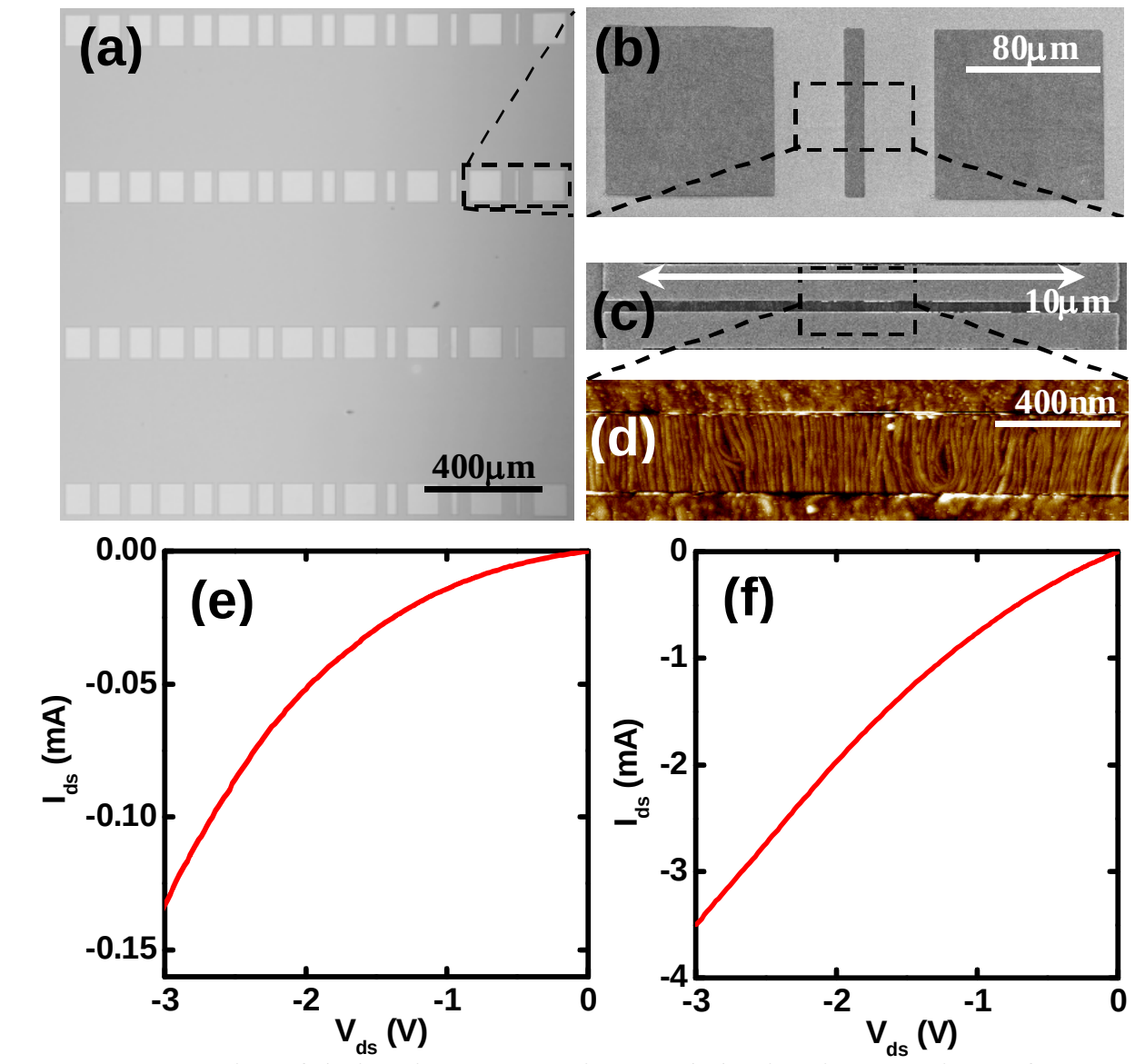

***Figure 3.*** Microfabrication patterning and device integration of SWNT LB films. (a) Optical image of a patterned SWNT LB film. The squares and rectangles are regions containing densely aligned SWNTs. Other areas are $SiO_2$ substrate regions. (b) SEM image of a region highlighted in (a) with packed SWNTs aligned vertically. (c) SEM image showing a 10-micron-wide SWNT LB film between source and drain electrodes formed in a region marked in (b). (d) AFM image of a region in (c) showing aligned SWNTs and the edges of the S and D electrodes. (e) Current vs. bias ($I_{ds}$-$V_{ds}$) curve of a device made of Hipco SWNTs (10μm channel width and 250nm channel length). (f) $I_{ds}$-$V_{ds}$ of a device made of laser-ablation SWNTs (10μm channel width and 250nm channel length).

The LB assembly of densely aligned SWNTs can be combined with chemical separation and selective chemical reaction methods[12] to afford purely metallic or semiconducting SWNTs in massive parallel configuration useful for interconnection or high speed transistor applications at large scale. The method is generic in terms of the type of nanotube materials and substrates.

**Acknowledgment.** We thank Dr. Pasha Nikolaev for providing laser-ablation SWNTs and MARCO-MSD and Intel for support.

**Supporting Information Available:** Experimental details are available free of charge via the internet at http://pubs.acs.org.

REFERENCES

1. Guo, J., Hasan, S., Javey, A., Bosman, G. & Lundstrom, M. *IEEE Trans. Nanotechnology* **2005**, 4, 715.
2. Zhang, Y., Chang, A. & Dai, H. J. *Appl. Phys. Lett.* **2001**, 79, 3155.
3. Huang, S. M., Maynor, B., Cai, X. Y. & Liu, J. *Adv. Mater.* **2003**, 15, 1651.
4. Kocabas, C., Hur, S., Gaur, A., Meitl, M. A., Shim, M. and Rogers, J. A. *Small* **2005**, 11, 1110.
5. Han, S., Liu, X. & Zhou, C. W. *J. Am. Chem. Soc.* **2005**, 127, 5294.
6. Gao, J., Yu, A., Itkis M. E., Bekyarova, E., Zhao, B., Niyogi, S. & Haddon, R. C. *J. Am. Chem. Soc.* **2004**, 126, 16698.
7. Rao, S. G., Huang, L., Setyawan, W. & Hong, S. *Nature* **2003**, 425, 36.
8. Guo, Y., Wu, J., Zhang, Y. *Chem. Phys. Lett.* **2002**, 362, 314.
9. Krstic, V., Duesberg, G. S., Muster, J., Burghard, M., Roth, S. *Chem. Mater.* **1998**, 10, 2338.
10. Star, A., Stoddart, J. F., Steuerman, D., Diehl, M., Boukai, A., Wong, E. W., Yang, X., Chung, S. W., Choi, H. & Heath, J. R. *Angew. Chem. Int. Ed.* **2001**, 40, 1721.
11. Kim, W. Javey, A., Tu, R., Cao, J., Wang, Q. & Dai, H. *Appl. Phys. Lett.* **2005**, 87, 173101.
12. Zhang, G. Y., Qi, P. F., Wang, X. R., Lu, Y. R., Li, X. L., Tu, R., Bangsaruntip, S., Mann, D., Zhang, L. & Dai, H. *Science* **2006**, 314, 974.

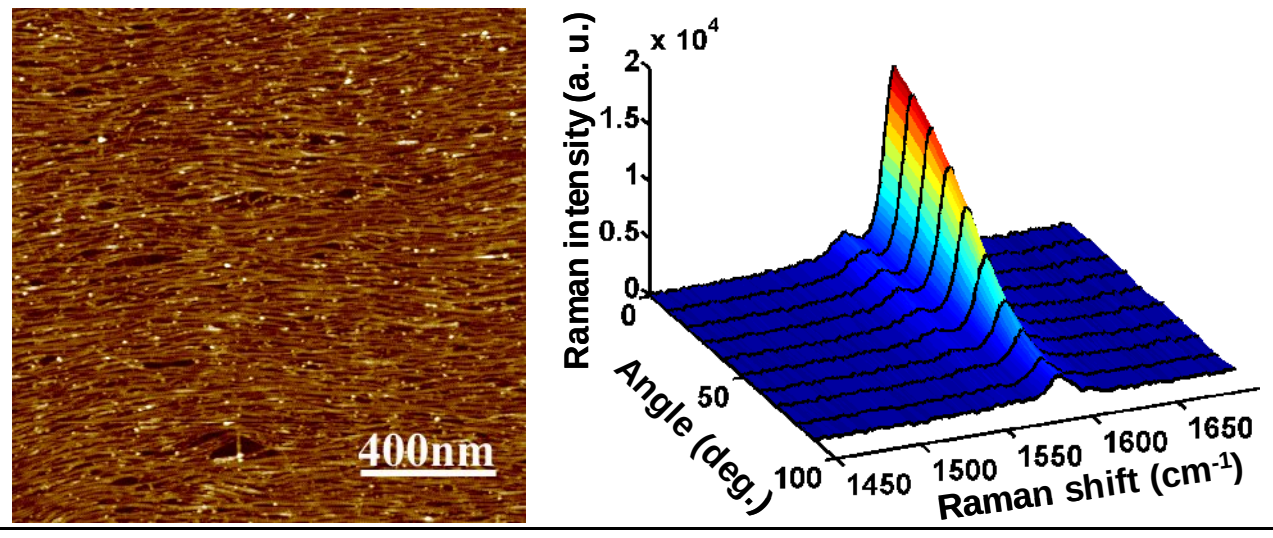

ABSTRACT FOR WEB PUBLICATION.

Single-walled carbon nanotubes (SWNTs) exhibit advanced electrical and surface properties useful for high performance nanoelectronics. Important to future manufacturing of nanotube circuits is large-scale assembly of SWNTs into aligned forms. Despite progress in assembly and oriented synthesis, pristine SWNTs in aligned and close-packed form remain elusive and needed for high-current, -speed and -density devices through collective operations of parallel SWNTs. Here, we develop a Langmuir-Blodgett (LB) method achieving monolayers of aligned SWNTs with dense packing, central to which is a non-covalent polymer functionalization by poly(m-phenylenevinylene-co-2,5-dioctoxy-p-phenylenevinylene) (PmPV) imparting high solubility and stability of SWNTs in an organic solvent 1,2-dichloroethane (DCE). Pressure cycling or 'annealing' during LB film compression reduces hysteresis and facilitates high-degree alignment and packing of SWNTs characterized by microscopy and polarized Raman spectroscopy. The monolayer SWNTs are readily patterned for device integration by microfabrication, enabling the highest currents (~3mA) through the narrowest regions packed with aligned SWNTs thus far.